\documentclass[prl,twocolumn, superscriptaddress]{revtex4}
\usepackage{graphicx,amsmath,amssymb,amsxtra}
\renewcommand{\narrowtext}{\begin{multicols}{2} \global\columnwidth20.5pc}

\def\be{\begin{eqnarray}}
\def\ee{\end{eqnarray}}

\newcommand{\Eq}[1]{Eq.~(\ref{#1})}
\newcommand{\Fig}[2]{Fig.~(\ref{#1}{#2})}

\newcommand{\ket}[1]{{|#1\rangle}}
\newcommand{\ketbar}[1]{{\overline{|#1\rangle}}}

\begin{document}

\title{Pfaffian statistics through adiabatic transport in the
1D coherent state representation}

\author{Alexander Seidel}
\affiliation{
Department of Physics, 
Washington University, St. Louis, MO 63136, USA}
\date{\today}

\begin{abstract}
Recent work has shown that the low energy sector of certain
quantum Hall states is adiabatically connected to simple 
charge-density-wave patterns that appear, e.g., when the system
is deformed into a thin torus. Here it is shown that the 
patterns emerging in this limit already determine the
non-abelian statistics of the $\nu=1$ Moore-Read state. Aside from the
knowledge of these patterns, the method
only relies on the principle of adiabatic continuity, 
the effectively noncommutative geometry in a strong magnetic field,
and topological as well as locality arguments. 
\end{abstract}

\maketitle


The discovery of the fractional quantum Hall (FQH)
effect \cite{TSG} has 
bestowed us with a fascinating problem.
Since then both theory
and experiment have demonstrated the vastness
of possible incarnations of FQH systems. While powerful methods 
have been developed to study
these states,
the objects under investigation have become ever more
complex. The current interest in 
quantum Hall physics
is greatly fueled by the exciting possibility
of non-abelian states\cite{mooreread,wenblok}. These 
may provide the ideal ``hardware'' for a
fault tolerant topological quantum computer\cite{kitaev,DFN}.
Recently, new descriptions of FQH systems have been
developed that make explicit use of the one-dimensional (1D)
Hilbert space structure of a Landau 
level\cite{seidel1, seidel2, karlhede, bernevig, seidel4}.
In this approach, simple 1D patterns are used to label the
ground states and quasi-particle/hole type excited states of
FQH systems. For both single-component\cite{seidel1, seidel2, karlhede, bernevig} and multi-component\cite{seidel4} FQH states, 
these 1D patterns naturally encode characteristic properties
of the underlying states, such as their torus degeneracies and the fractional
charges of elementary excitations. 
Their equivalence to "patterns of zeros" was clarified
in Ref. \onlinecite{wenwang}, which also introduced a framework to
construct new quantum Hall states. 
A fundamental problem in quantum Hall physics
is the calculation of braiding statistics
through adiabatic transport of quasi-particles. 
The classical
treatment of Ref. \onlinecite{ASW} has proven 
exceedingly difficult to generalize to
non-abelian states. For abelian states,
an alternative
treatment was recently given in
Ref. \onlinecite{seidel3}, where quasi-hole
states were identified as
coherent states formed in the basis defined by the 1D 
patterns. 
Here, I will show that 
a refined version
of this approach is capable of 
extracting the
non-abelian statistics of quasi-holes in Moore-Read (Pfaffian) states
\cite{mooreread} as well. 
These quasi-holes have
been argued to carry gapless Majorana fermions,
based on the conformal field theory (CFT) construction
of these states\cite{mooreread, naywil}, and the analogy between
the Pfaffian and a $p+ip$ wave superconductor\cite{RG,ivanov,stern,oshikawa}.
This is confirmed by the
present approach,  even
though Majorana fermions are in no way basic ingredients
included into the description of Pfaffian states presented here.\\
\indent{\em Pfaffian quasi-holes as coherent states.}
For simplicity, I will focus on the bosonic Pfaffian at 
filling factor $\nu\!=\!1$,
on toroidal topology.
The torus will be identified with
a rectangular two-dimensional
(2D) domain of dimensions $L_x$ and $L_y$ subject to
periodic (magnetic) boundary conditions. I will set the
magnetic length equal to $1$, such that $L_xL_y=2\pi L$,
where L is the number of magnetic flux quanta 
through the surface, or the number of orbitals in the
lowest Landau level (LLL).
One way to uniquely attach distinct 1D patterns
to the degenerate Pfaffian ground states is the following.
Consider the local Hamiltonian $H_{Pf}$ \cite{GWWprl}
whose unique ground states are the $\nu=1$ Pfaffian 
wavefunctions.
One can observe that in the ``thin'' torus limit $L_y\!\ll\! 1$,
$L_x\!\gg\!1$,
the ground states of $H_{Pf}$ describe charge-density-wave 
(CDW) states, which are adiabatically connected to the
bulk quantum Hall states in the 2D limit $L_x,L_y\!\gg\!1$\cite{seidel2}.
These 1D CDW patterns run along the $x$ direction.
The underlying states are trivial basis states of the 
LLL Fock space, where the orbitals $\varphi_n$ of a
certain LLL basis have definite occupation numbers.
For consecutive orbitals $\varphi_n$, these occupation numbers
form periodic patterns, such as $20202020\dotsc$,
$02020202\dotsc$ and $11111111\dotsc$ for the three
Pfaffian ground states at $\nu=1$ on the torus\cite{seidel2}.
From the get-go, it should be noted that there is a
notion of ``S-duality'' in this formalism, which 
comes about by interchanging the roles of $x$ and $y$
(i.e. through modular S-transformations on the torus):
In the opposite thin torus limit, 
$L_x\!\ll\! 1$,
$L_y\!\gg\!1$, the ground states must evolve
into
the same CDW patterns, now extending along $y$. These
states
are then occupation number eigenstates in a 
LLL basis $\bar\varphi_n$, which can be thought
of as a ``rotated'' version of the basis $\varphi_n$.
In general, a ground state that evolves into a given
1D pattern in one thin torus limit will evolve into
a superposition of such patterns in the opposite
thin torus limit, and vice versa.
One can argue\cite{seidel2,karlhede} that the adiabatic continuity
is observed not only by the incompressible ground states,
but also by states with $2n$ quasi-hole type excitations.
In the thin torus limit, such states evolve into
CDW patterns which are domain walls between
ground state patterns of the $11$-type, and of the
$02$- or $20$-type, respectively, e.g. $202020111110202020$.
These patterns are directly related 
to the 
fractional charge 
$\nu/2$
of the Pfaffian quasi holes, and
to the number of topological sectors for $2n$-hole 
states\cite{seidel2, karlhede}. Let us consider a pattern
with two domain walls whose positions
along the $x$-axis are $x_1$, $x_2$, e.g. as shown
\Fig{fig1}{a}. 
We can define orbital positions $a_1=x_1/\kappa$ and $a_2=x_2/\kappa$
of the domain walls,
where
$\kappa=2\pi/L_y$ is the distance between consecutive
orbitals in the LLL basis $\varphi_n$. 
Roughly, $a_1$ and $a_2$
should correspond 
to the zeros adjacent to the $11$-string 
in the pattern.
There is
some freedom, however, in the precise definition of
the domain wall position
relative to the end of a $11$-string,
which I will get back to below. Given some convention to define
$a_1$ and $a_2$, I will denote the $L_y\rightarrow 0$
CDW state corresponding to a given pattern by $|a_1,a_2,c)$.
Here, $c$ labels one of the four topological sectors for 2-hole states
on the torus depicted in \Fig{fig1}{a}. These sectors
differ
by the order of the $11$- and $20$-type strings
relative to the domain walls, and by the two possible 
phases of the $20$-type CDW patterns.
The state that
is obtained from $|a_1,a_2,c)$ by slowly 
increasing the circumference $L_y$ is denoted by
$|a_1,a_2,c\rangle=\hat S (L_y)|a_1,a_2,c)$, where
$\hat S(L_y)$ is a unitary operator that describes the
adiabatic evolution from $L_y=0$ to finite $L_y$.
Note that as long as adiabatic continuity holds,
the states $|a_1,a_2,c\rangle$ must be orthogonal
for different $a_1$, $a_2$, $c$, since the $|a_1,a_2,c)$
are. $|a_1,a_2,c\rangle$ is now a state with two 
quasi-holes that are
localized in $x$ at $\kappa a_1$ and $\kappa a_2$,
but are entirely delocalized around the $y$ circumference.
It must be stressed that for $L_x$, $L_y$ both large,
the state $\ket{a_1,a_2,c}$ has a constant 
density profile away from $x=\kappa a_1,\kappa a_2$,
and no CDW pattern is visible\cite{seidel2}.
In this 2D regime,
 consider now a state with two holes 
in the sector $c$, that are
localized at complex
coordinates $h_j=h_{jx}+ih_{jy}$, $j=1,2$,
and that are well separated along the $x$ axis, $h_{2x}\!-\!h_{1x}\gg 1$.
For such states, the following coherent 
state expression
can be given in terms of the $\ket{a_1,a_2,c}$:
\vspace{-2mm}
\begin{subequations}\label{coherent}
\begin{align}
& \ket{\psi_c(h_1,h_2)}={\cal N}\sideset{}{'}\sum_{a_1<a_2} \phi(h_1, \kappa a_1)
\phi(h_2, \kappa a_2) \ket{a_1 a_2 c}\label{coherenta}\\
&\text{where}\; \phi(h,x)=\exp({\frac{i}{2}(h_y+\delta/\kappa)x-\frac{1}{4}(x-h_x)^2}),\label{phi}
\vspace{-2mm}
\end{align}
\end{subequations}
and $\cal N$ is a constant.
An expression of this form was derived in Ref.
\onlinecite{seidel3} for Laughlin states by making contact
with wavefunctions. However, it was also argued there that
the general form of \Eq{coherent} essentially follows
from the effective noncommutative geometry of the physics
in the LLL, where a commutation relation of the form
$[\hat x, \hat y]\propto i$ holds. The prime
in \Eq{coherent} indicates that the $a_i$ are restricted
to be of the form $a_i\!=\!2 n_i\! +\!f_i(c)$, $n_i$ integer,
and $f_i(c)$ is defined such that $a_i$ is an allowed 
domain wall position for the $i$th domain wall in the 
sector $c$, see below. Within a given sector, we can
raise the domain wall position $a_i$ only in units 
of $2$. With this, 
$\phi(h, \kappa a_i)$ can also be viewed
as a LLL wavefunction (in $h$) 
of a charge $1/2$ particle \cite{seidel3,note2},
with boundary conditions in $h_y$ that are sensitive
to the sector $c$ and to $i$.
\Eq{coherent} also captures the fact that the
$y$-position of a hole enters as $x$-momentum,
as required by $[\hat x, \hat y]\propto i$. However,
we can know the relation between $y$-position and $x$-momentum
only up to a constant $\delta$ at this point, which will be
further specified below. The goal of this paper is to 
determine whether the simple form of \Eq{coherent} can
capture even non-abelian statistics. To this end,
let us observe that \Eq{coherent} could easily be
generalized to $2n$ holes, where we must keep in mind
that we know the precise form of the coherent state expression only
when the holes are well separated along the $x$ axis.
Let us consider 4 holes, in the sector that is represented
by the pattern shown in  \Fig{fig1}{b}. 
Suppose we adiabatically exchange holes $2$ and
$3$ in \Fig{fig1}{b}. As long as the holes are well separated in $x$,
\Eq{coherent} does not permit a transition
into a different topological sector. Such
a transition would imply a shift of the phase
of the $20$-CDW in the corresponding sector
label, as shown in \Fig{fig1}{b}. It is suggestive that
such transitions should not happen while all
$20$-strings are ``long'', as this would apparently
affect many microscopic degrees of freedom at a time.
Such interesting transitions should only happen
when the two holes are close in $x$ (though not in $y$),
and \Eq{coherent} is not valid. By the same argument,
an exchange of holes $1$ and $2$ or $2$ and $3$
should not change the topological sector at all.
These holes are linked by a $11$-type string in the sector
label, and all $20$-strings remain long during their exchange,
implying topological stability. 
These considerations
immediately tell us that 
there is a basis of $2n$-hole states
on the torus, for which every second generator of the braid group
is diagonal, and every {{\em other} generator is block diagonal
with block size 2. 
However, we do not yet know anything about
complex phases picked up during the exchange processes,
or amplitudes for the allowed transitions. I will now
discuss how such information can be obtained using methods
similar to those
of Ref. \onlinecite{seidel3}.\\
\indent{\em Locality and duality constraints.}
The result of exchanging two holes
should affect the thin torus limit of
the state only for orbitals $\varphi_n$
that participated in the exchange,
as shown in \Fig{fig1}{b}. This follows from
the locality of $H_{Pf}$ and the short
range of the correlations in the Pfaffian
ground states.
Consider a state of $2n$ holes that are well separated
along the $x$-axis. The exchange of
two adjacent holes should thus act on the sector
label of the state only by affecting the CDW-string
linking the associated domain walls.
We can thus work out the
transformation on the sector label in a $2n$-hole
state, as long as we know how 2-hole states
transform under exchange
for all sectors shown in
\Fig{fig1}{a}. 
However, we must study 2 holes on a torus
for both even and odd particle number $N=L\!-\!1$, since
it will in general make a difference whether the
$11$-string linking two holes is even or odd in length. 
To overcome the technical difficulty that \Eq{coherent}
is sensible only for $h_{2x}\!-\!h_{1x}\gg 1$,
we could have worked from the opposite, or ``dual'',
thin torus limit where $L_x\rightarrow 0$. 
There must be an expression analogous to \Eq{coherent}
in terms of adiabatically continued domain wall states
from this dual thin torus limit, 
which I denote by $\ketbar{a_1,a_2,\bar c}$:
\vspace{-1mm}
\begin{equation}\label{dual}
\ketbar{\psi_{\bar c}(h_1,h_2)}={\bar {\cal N}}\sideset{}{'}\sum_{a_1<a_2} \bar\phi(h_1, \bar\kappa a_1)
\bar\phi(h_2, \bar\kappa a_2) \ketbar{a_1 a_2 \bar c}
\vspace{-1mm}
\end{equation}
where $\bar\kappa=2\pi/L_x$, $\bar\phi(h,y)=\phi(-ih, y)$, and \Eq{dual} is now valid for $h_{2y}-h_{1y}\gg 1$.
The sector label $\bar c$ refers to the
same four patterns depicted in \Fig{fig1}{a},
which, however, extend along the
$y$ axis in the $L_x\rightarrow 0 $ limit. 
While the states 
$\ket{\psi_c(h_1,h_2)}$, $\ketbar{\psi_{\bar c}(h_1,h_2)}$ 
should span the same subspace at given $h_1$, $h_2$, 
their precise relation is not immediately obvious. 
Nonetheless, on each segment of the exchange path depicted in \Fig{fig1}{c}, at least one of the two coherent state expressions Eqs. \eqref{coherent}, 
\eqref{dual} is 
valid. If we knew the transition matrices $u_{c\bar c}(h_1,h_2)$ that describe the
change of basis between $\ket{\psi_c}$ and $\ketbar{\psi_{\bar c}}$ in regions 
where both expressions are valid, we could calculate the adiabatic transport for the
exchange path shown in \Fig{fig1}{c}. To make progress, we need to specify the 
quantities
$f_i(c)$ introduced via $a_i\!=\!2 n_i\! +\!f_i(c)$ above. 
Let the position of the first
(last) zero of a thin torus pattern
in the sector $c=1$ be $2n_{1}$ ($2n_{2}$), see \Fig{fig1}{a}.
We define the corresponding domain
wall positions to be $a_{1,2}=2n_{1,2}\mp s$.
That is, we let  $f_1(1)=-s$, $f_2(1)=s$, 
where $s$ is a parameter to be determined from locality,
see below.
Similarly, for $c=3$ we must then let 
$f_1(3)=s-1$, $f_2(3)=\eta-s$, 
where $\eta\!=\!0$ ($\eta\!=\!1$) for $N$ even (odd). 
The sectors $c\!=\!2$, $4$ are related to  $c\!=\!1,3$ by 
translation, 
i.e. $f_i(2)=f_i(1)+1$, $f_i(4)=f_i(3)+1$. 
We can now analyze the properties
of the states Eqs. \eqref{coherent}, 
\eqref{dual} under magnetic translations,
$T_x$, $T_y$. Since $T_{x,y}$ commute with the 
adiabatic
evolution operator,
the states
$\ket{a_1,a_2,c}$, $\ketbar{a_1,a_2,c}$,
 transform
under translations in the same way as the trivial CDW states $|a_1,a_2,c)$,
$\overline{|a_1,a_2,c)}$. The action of $T_{x,y}$
on such states is discussed in detail, e.g., in Ref. \onlinecite{seidel3}.
For large $L_x$, $L_y$, it turns out that \Eq{coherenta} is almost an eigenstate
of $T_y$, while \Eq{dual} is almost an eigenstate of $T_x$, up to terms
of order $1/\sqrt{L}$. Since close to $\nu=1$, 
$T_x$ and $T_y$ almost commute,
we can form linear 
combinations of the states \Eq{coherenta} that
are almost eigenstates of both $T_x$ and $T_y$:
\vspace{-2mm}
\begin{equation}
\begin{split}\label{trans}
  T_x \ket{\psi_{\mu,\nu}(h_1,h_2)}&\simeq \mu e^{-i\kappa(h_{1y}+h_{2y})/2+i\pi\eta}
\ket{\psi_{\mu,\nu}(h_1,h_2)}\\
T_y \ket{\psi_{\mu,\nu}(h_1,h_2)}&\simeq \nu e^{i\kappa(h_{1x}+h_{2x})/2+i\pi\eta}
\ket{\psi_{\mu,\nu}(h_1,h_2)}
\end{split}
\vspace{-2mm}
\end{equation}
Here, $\mu,\nu=\pm 1$, and $\simeq$ means equality up to terms that
vanish when $L\rightarrow\infty$. 
Likewise,
we must be able to form superpositions $\ketbar{\psi_{\mu,\nu}(h_1,h_2)}$ of the
states \Eq{dual} that have {\em the same} properties under translations
as the $\ket{\psi_{\mu,\nu}(h_1,h_2)}$. 
Consistency then requires $\delta=0$ or $\delta=\pi$\cite{note1}. Omitting the
$h_{1,2}$ dependence, one finds
\begin{figure}[t]
\begin{center}
\includegraphics[width=7.9cm]{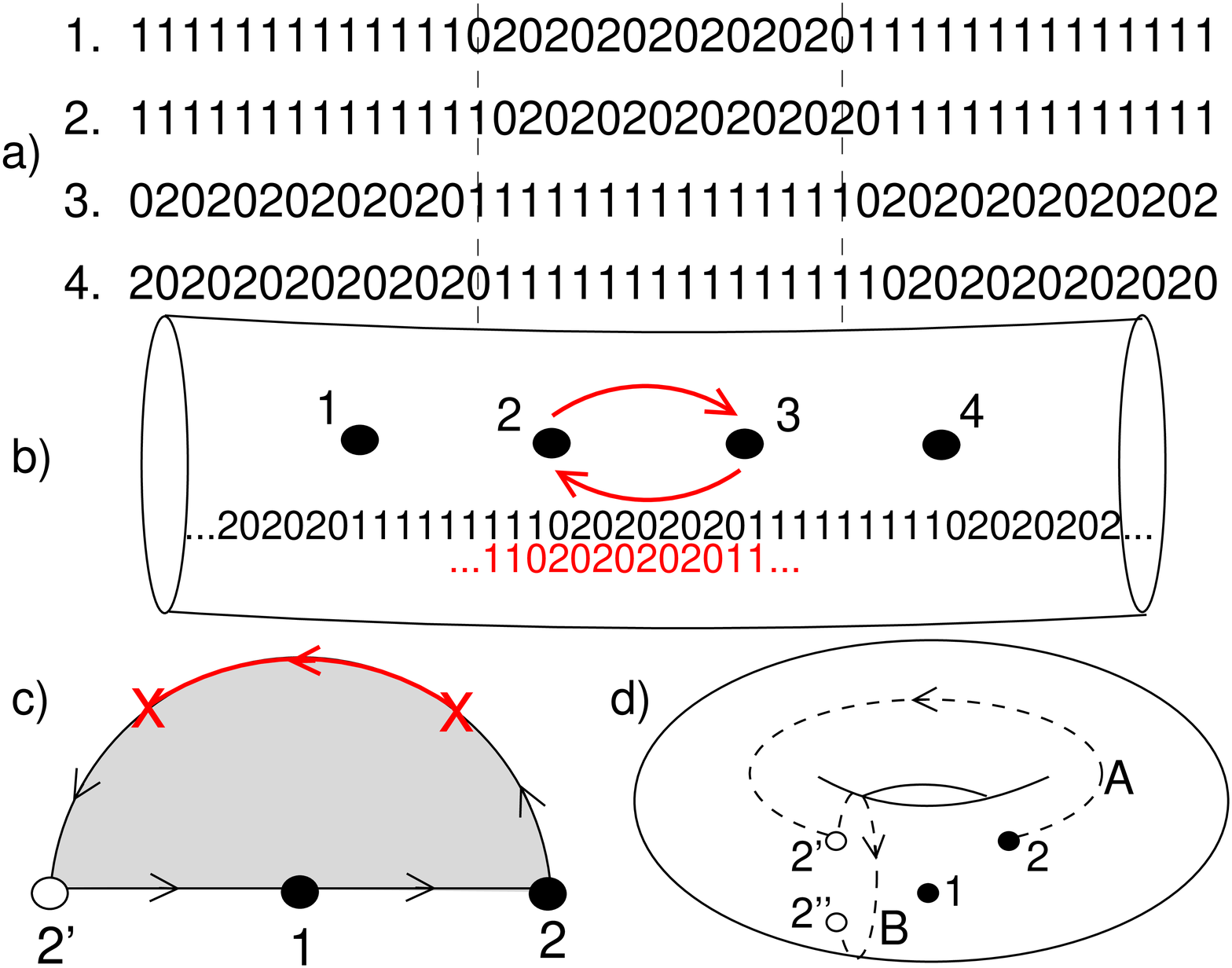}
\end{center}
\caption{a) Thin torus patterns defining 4 sectors for 2-hole states. 
Vertical lines mark the orbitals $2n_1$, $2n_2$, see text. 
b) State of 4 holes.
A transition
into a different 
sector through exchange
of holes 2 and 3 is indicated. c) Exchange path for two holes. First, hole
2 is dragged along the arc into 2'. Then 2' and 1 are shifted horizontally.
At the points marked ``X'', the coherent state representations Eqs. 
\eqref{coherent}, \eqref{dual} 
are replaced by one another via 
\Eq{u}.
d) Paths A and B used to link the transition functions in \Eq{u} in different sectors. 
2' and 2'' always denote intermediate or final 
positions of hole 2.\label{fig1}}
\end{figure}
\begin{equation}
\begin{split}\label{munu}
   \ket{\psi_{\mu,\nu}}&=(\ket{\psi_{c=2-\nu}}+\mu e^{i\delta+i\pi\eta}\ket{\psi_{c=3-\nu}})\,/\sqrt{2}\\
 \ketbar{\psi_{\mu,\nu}}&=(\ketbar{\psi_{c=2-\mu}}+\nu e^{i\delta+i\pi\eta}\ketbar{\psi_{c=3-\mu}})\,/\sqrt{2}.
\end{split}
\end{equation} 
The $\ketbar{\psi_{\mu\nu}}$ then also satisfy \Eq{trans}. For the
two bases $\ket{\psi_{\mu\nu}}$, $\ketbar{\psi_{\mu\nu}}$, the transition
matrix must thus be diagonal. That is, for $h_{2x}\!-\!h_{1x}\!\gg\! 1$,
$|h_{2y}\!-\!h_{1y}|\! \gg\! 1$, we must have:
\begin{equation}
\begin{split}\label{u}
  \!\! \ket{\psi_{\mu\nu}(h_1,h_2)} &= u^+_{\mu\nu}(h_1,h_2) \ketbar{\psi_{\mu\nu}(h_1, h_2)} \; (h_{2y}>h_{1y})\\
\!\! \ket{\psi_{\mu\nu}(h_1,h_2)} &= u^-_{\mu\nu}(h_1,h_2) \ketbar{\psi_{\mu\nu}(h_2, h_1)}
\; (h_{2y}<h_{1y})
\end{split}
\end{equation}
where $u^\pm_{\mu\nu}(h_1,h_2)$ is a pure phase. The process of
exchanging the two holes is then diagonal in the $\mu,\nu$-basis.
From Eqs. \eqref{coherent}, \eqref{dual}, we can easily calculate local
Berry connections in the limit $\kappa, \bar\kappa\ll 1$ (large system)
via
\vspace{-1mm}
\begin{equation}
\begin{split}\label{berry}
 i\langle \psi_{\mu\nu}|\nabla_{h_{1,2}}|\psi_{\mu\nu}\rangle
&=-\frac 12\, (0 , h_{1,2;x})\\
i\langle\overline{\psi_{\mu\nu}|}\nabla_{h_{1,2}}
\overline{|\psi_{\mu\nu}\rangle}
&=\;\;\,\frac 12 \,(h_{1,2;y},0).
\end{split}
\end{equation}
From \Eq{berry}, we can locally determine the $h_{1,2}$
dependence of $u^\pm_{\mu\nu}$ up to an overall phase,
for given $\mu$, $\nu$ and for each ``region''
$\sigma\equiv\text{sgn}(h_{2y}-h_{1y})$:
\begin{equation}\label{alpha}
 u^\sigma_{\mu\nu}(h_1,h_2)=e^{i\alpha^\sigma_{\mu\nu}}e^{i(h_{1x}h_{1y}+h_{2x}h_{2y})/2}\,.
\end{equation}
The Berry phase $\beta_{\mu\nu}$ for the exchange
of two holes in the state
$\ket{\psi}_{\mu\nu}$ along the path
 shown in \Fig{fig1}{c}
can now be calculated from Eqs. \eqref{u}--\eqref{alpha},
as shown in 
detail in  \cite{seidel3}:
\begin{equation}
\label{gamma}
\beta_{\mu\nu}=\Phi_{AB} + \alpha_{\mu\nu}^+-\alpha_{\mu\nu}^-\;.
\end{equation}
Here, the Aharonov-Bohm phase $\Phi_{AB}$ 
equals $-1/2$ times the flux through the shaded
area in \Fig{fig1}{c}. I denote the statistical 
part
of the Berry phase by 
$\gamma_{\mu\nu}=\beta_{\mu\nu}-\Phi_{AB}$.
To further determine $\gamma_{\mu\nu}$, we note
that the functions $u_{\mu\nu}^\sigma$
defined in \Eq{u} 
for different sectors $\mu$, $\nu$
and $\sigma$
come in two classes. For given $\sigma\mu\nu=\pm 1$,
these functions can be related to each other
by moving one hole along one of the paths
A or B
shown in \Fig{fig1}{d}. Both Eqs. \eqref{coherent}, \eqref{dual}
are valid along these paths, which rearrange
the hole positions such that $\sigma\rightarrow -\sigma$.
At the same time, it is found that $\nu\rightarrow -\nu$
for path A and  $\mu\rightarrow -\mu$
for path B. Related to that, along path A
the sector $c$ changes to $c'=5-c$ for
$2\eta\!+\!c\!-\!3\!\ge\!0$, $c'=c+2$ otherwise.
We may identify $\ket{a_1,a_2,c}\equiv\ket{a_2-L,a_1,c'}$,
and similarly $\ketbar{a_1,a_2,\bar c}\equiv\ketbar{a_2-L,a_1,\bar c'}$.
The phases picked up by the states \Eq{munu}
along paths A,B follow from these
identifications, the
non-single-valuedness of Eqs. \eqref{coherent}, \eqref{dual}
in $h_{1,2}$,  
and the Berry connection \Eq{berry}.
This leads to relations among the $\alpha^\sigma_{\mu\nu}$.
As a result, one can express
all the statistical phases $\gamma_{\mu\nu}$ through
$\gamma_{++}$ and $s$,
for both even ($e$) and odd ($o$)
particle number. It is found that 
$\gamma_{-+}^e=\gamma_{+-}^e=-\gamma_{++}^e+2\pi s$, 
$\gamma_{--}^e=\gamma_{++}^e+\pi-4\pi s$ for $N$
even, while $\gamma_{-+}^o=\gamma_{+-}^o=-\gamma_{++}^o+2\pi s$, 
$\gamma_{--}^o=\gamma_{++}^o-4\pi s$ for $N$
odd. Finally, let us consider the locality
constraints on the result of exchanging two holes,
as discussed initially. If the holes are in 
the sector $c=1$ or $c=2$, the $11$-part
of the thin torus limit of the states $\ket{a_1,a_2,c}$
entering \Eq{coherent} will be arbitrarily long
in the thermodynamic limit, during the entire
exchange process. The parity of its length, which
is the parity of $N$, should thus have no influence
on the exchange. 
In terms of the $\mu\nu$-basis \Eq{munu},
this immediately implies $e^{i\gamma_{++}^e}=e^{i\gamma_{-+}^o}$,
$e^{i\gamma_{-+}^e}=e^{i\gamma_{++}^o}$. 
On the other hand, in the sectors $c=3$ and $c=4$,
the result of the exchange may depend on
the particle number parity. However, as argued
above, in both these sectors the exchange should
result in a pure phase, which must be the same for
$c=3$ and $c=4$ by translational symmetry. This
leads to $e^{i\gamma_{--}^{e,o}}=e^{i\gamma_{+-}^{e,o}}$.
These requirements severely constrain the parameters 
$s$, $\gamma_{++}^{e,o}$. Solutions are
of the form
\begin{subequations}\label{result}
\begin{align}
 \gamma^e_{++}&=\gamma^o_{+-}=\gamma^o_{-+}=\gamma^o_{--}=5\pi/8-3\pi r/4\label{odd}\\
\gamma^o_{++}&=\gamma^e_{+-}=\gamma^e_{-+}=\gamma^e_{--}=\pi/8+\pi r/4\,,\label{even}
\end{align}
\end{subequations}
where $r$ is integer. For $r=0$ the
two possible statistical phases in the even/odd
particle number sector are $\pi/8$ and $5\pi/8$,
which coincides with the prediction based on CFT
methods\cite{nayak08}. 
In terms of the original sectors displayed in \Fig{fig1}{a},
\Eq{result} means the following: When exchanging
two holes that are linked by a $11$-string in the sector
label, the state picks up a phase given by \Eq{odd}
if the $11$-string has odd length, and by \Eq{even}
if it has even length. If the holes are linked
by a $20$-type string, it follows from Eqs.
\eqref{munu}, \eqref{result} that the final
state remains in the original sector with
an amplitude $e^{i\theta}/\sqrt{2}$, and transitions
into the sector where the $20$-string has the opposite phase
with an amplitude $(-1)^rie^{i\theta}/\sqrt{2}$.
Here, $\theta/\pi=1/4+(-1)^r/8+\lfloor r/2\rfloor/2$.
When applied to a general $2n$-hole label as depicted in
\Fig{fig1}{b}, this can be shown to be equivalent to the following
representation of the generator that exchanges holes
$j$ and $j+1$:
\vspace{-1mm}
\begin{equation}\label{majo}
e^{i\theta} e^{(-1)^r \frac \pi 4\, \eta_j \eta_{j+1} },
\vspace{-2mm}
\end{equation}
where $\eta_k$ is a Majorana fermion associated
with the $k$-th domain wall in the label. Here, any sector
label denotes an eigenstate of $\eta_{2j-1}\eta_{2j}$
with eigenvalue $+i$ ($-i$), if the associated
domain walls are linked by a $11$-string of odd (even)
length. Since the $(-1)^r$ in \Eq{majo} can be absorbed
by a unitary transformation, this determines the
non-abelian part of the statistics unambiguously.\\
\indent{\em Conclusion.}
It was shown that through the 
general ansatz \Eq{coherent}, the thin torus
CDW patterns associated with the $\nu=1$ Pfaffian already
determine its non-abelian statistics.
The result \Eq{majo}
agrees with the Majorana fermion picture
based on conformal field theory (CFT) and the $p\!+\!ip$-wave
superconductor analogy. The abelian part of the
statistics is also determined modulo $\pi/4$. 
This serves as an independent check 
for the statistics derived from
CFT methods, which have so far not been proven via
adiabatic transport. 
A generalization of the
formalism presented here to more complicated non-abelian
states seems natural.\\
\indent{\em Note:} After this work had been
submitted, a preprint by Read appeared
\cite{readpreprint}
discussing adiabatic transport based on
conformal block wavefunctions.  

\begin{acknowledgments}
I would like to thank N. Bonesteel, K. Yang, and D.-H. Lee
for stimulating discussions.
\end{acknowledgments}

\vspace{-4mm}

\begin{thebibliography}{23}
\bibitem[{not({\natexlab{b}})}]{note1}
\bibinfo{note}{{This was already assumed in \Eq{trans}. }}.
\end{thebibliography}

\end{document}